\documentclass[12pt]{article}
\usepackage{amsmath}
\usepackage{graphicx}
\usepackage{amssymb}
\usepackage{amsfonts}
\usepackage{color}
\usepackage{indentfirst}

\def\beq{\begin{eqnarray}}
\def\eeq{\end{eqnarray}}
\def\nn{\nonumber}                %%%    nonumber
\def\ln{\,\mbox{ln}\,}
\def\Ln{\,\mbox{Ln}\,}

\def\Tr{\,\mbox{Tr}\,}

\def\al{\alpha}
\def\be{\beta}

\def\ga{\gamma}
\def\de{\delta}
\def\vp{\varepsilon}
\def\ep{\epsilon}

\def\ka{\kappa}
\def\la{\lambda}
\def\na{\nabla}
\def\pa{\partial}

\def\si{\sigma}

\def\ph{\varphi}
\def\ta{\tau}

\def\Ga{\Gamma}
\def\De{\Delta}

\DeclareMathOperator{\cx}{\square}

\begin{document}

%%%%%%%%%%%%%%%%%%%   Wagno's set:
% \usepackage{correctmathalign} % Correction Spacings: eqnarray(*), aligned, alignedat, and gathered
	
\begin{center}
%%%%%%%%%%%%%%%%%%%%%%%%%%%%%%%%%%
\renewcommand*{\thefootnote}{\fnsymbol{footnote}} % redefining the output of the footnote counter

{\Large
Trace anomaly and induced action for a metric-scalar background}
%% interaction term (draft)}
\vskip 6mm
		
{Manuel Asorey}$^{a}$
\hspace{-1mm}\footnote{E-mail address: \ asorey@unizar.es},
%%%%%%%%%%%%%%%%%%%%%%%%%%%%%%%%%%%%%%
\ \ {Wagno Cesar e Silva}$^{b}$
\hspace{-1mm}\footnote{E-mail address: \ wagnorion@gmail.com},
%%%%%%%%%%%%%%%%%%%%%%%%%%%%%%%%%%%%%%
\ \ {Ilya L. Shapiro}$^{b}$
\hspace{-1mm}\footnote{E-mail address: \ ilyashapiro2003@ufjf.br},
%%%%%%%%%%%%%%%%%%%%%%%%%%%%%%%%%%%%%%
\ \ {P\'ublio R. B. do Vale}$^{b}$
\hspace{-1mm}\footnote{E-mail address: \ publiovale@gmail.com}
\vskip 6mm
		
${a)}$ Centro de Astropart\'{\i}culas y F\'{\i}sica de Altas Energ\'{\i}as,
Departamento de F\'{\i}sica Te\'orica,
Universidad de Zaragoza, E-50009 Zaragoza, Spain
\vskip 2mm

${b)}$ Departamento de F\'{\i}sica, ICE, Universidade Federal de Juiz de Fora,
\\
Juiz de Fora, 36036-900, Minas Gerais, Brazil
\end{center}
%%%%%%%%%%%%%%%%%%%%%%%%%%%%%%%%%%
\vskip 2mm
\vskip 2mm
	
%\begin{history}
%\received{Day Month Year}
%\revised{Day Month Year}
%\end{history}
	
%%%%%%%%%%%%%%%%%%%%%%%%%%%%%%%%
\begin{abstract}
		
\noindent
The conformal anomaly and anomaly-induced effective action represent
useful and economic ways to describe semiclassical contributions to
the action of gravity. We discuss the anomaly in the case when the
background is formed by metric and scalar fields and formulate the
induced action in two standard covariant forms. The analysis of
induced action at low energies reveals existing connection to the
renormalization group and effective potential. The classification of
anomalous terms is extended to the scalar background and ambiguities
in the total derivative terms in the anomaly are considered using
Pauli-Villars regularization.
\vskip 3mm
		
		
%% \noindent
%% \textit{MSC:} 81T50, 81T20, 81T15,
%% 83C45  	Quantization of the gravitational field
%%%%%%%%%%%%%%%%%%%%%%%%%%%%%%%%
\end{abstract}
	
%\ccode{PACS numbers:}
%\tableofcontents
\setcounter{footnote}{0} % to reset the counter after such a change
\renewcommand*{\thefootnote}{\arabic{footnote}} % to switch back to Arabic numbering
%%%%%%%%%%%%%%%%%%%%%%%%%%%%%%%%
%%%%%%%%%%%%%%%%%%%%%%%%%%%%%%%%
%%%%%%%%%%%%%%%%%%%%%%%%%%%%%%%%
\section{Introduction}
\label{sec1}
	
The conformal anomaly \cite{CapDuf-74,duff77,ddi,duff94} and
anomaly-induced effective action \cite{rie,frts84} (see also
\cite{AntMot92} and \cite{PoImpo} for review
and further references) play an important role in the description
of loop corrections in the semiclassical approach. Even in the context
of the Standard Model (SM) one can use the conformal anomaly at
high energies, which can be still much below the Planck scale where
the full quantum gravity is supposed to become relevant. For example,
energy range of inflation, of the order $10^{11}-10^{14}$ GeV form
a scenario when the conformal anomaly may apply perfectly well
since the masses of the SM particles is negligible while the Planck
energies are far beyond.
	
One of the fundamental features of the Standard Model of weak
interactions is the Higgs mechanism of providing masses to $W$ and
$Z$ bosons and fermions.
%%  red
%%\textcolor{red}{
The use of conformal symmetry was traditionally invoked to
improve on the naturalness of the Standard Model (see, e.g.,
Refs. \cite{Bardeen-1995} and \cite{Meissner-2007}).
On the other hand, there is a growing interest in
the quantum effects on Higgs scalar, related to the vacuum
instability at high energies (in the UV), including the effects of
curved spacetime \cite{Higgsmass12,PlanckH12,stabV12}. Although
several mechanisms have been proposed to resolve this issue, the
subject does not look completely clear. Thus, it may be interesting
to consider in detail the quantum contributions to the scalar action
in the UV, using the powerful formalism of integrating conformal
anomaly.

The anomaly in a metric-scalar theory, including self-interaction in
the scalar sector, has been well discussed in the literature using
different methods \cite{DrumSho79,Hathrell,ValGoni87,AlvBarc}
(see also references therein), however we hope to add several
relevant aspects to the subject. In what follows, we derive and
discuss the trace anomaly and the anomaly-induced effective action
for the case when the background is formed by metric and also by
scalar fields.  Our considerations cover the limits of consistency
of the approach based on anomaly, that includes the discussion of
ambiguities in the anomaly and  induced action.

Last, but not least, we show how to evaluate the anomaly-based
contributions at low energies. Indeed, taking the IR limit in a
massless theory is a non-trivial issue. In our opinion, a better
understanding of the corresponding approximation and its
relation to other approaches look interesting
by themselves.

The paper is organized as follows. In Sec.~\ref{sec2} we describe
the anomaly for arbitrary particle contents of the semiclassical
theory. Sec.~\ref{sec4} discusses the anomaly-induced effective
action of metric and scalar field.  Sec.~\ref{sec3} is devoted to
the analysis of ambiguity in the anomaly, extending the results of
the previous works on the subject \cite{anomaly-2004,BoxAno}.
In Sec.~\ref{sec5}, we consider the low-energy limit of induced
action and the connection with renormalization group and effective
potential of scalar fields. Finally, in Sec.~\ref{sec6} we draw our
conclusions and discuss a few open issues, especially those related
to conformal quantum gravity.
%%   red
Throughout the paper, we use
pseudo-Euclidean notations, regardless of the Wick rotation to
Euclidean signature space is assumed when we refer to the
heat kernel results.

%%%%%%%%%%%%%%%%%%%%%%%%%%%%%%%%
%%%%%%%%%%%%%%%%%%%%%%%%%%%%%%%%
%%%%%%%%%%%%%%%%%%%%%%%%%%%%%%%%
\section{Conformal anomaly with scalar fields}
\label{sec2}

Consider first the toy model with $N$ copies of Dirac spinor fields
${\Psi}_k$ and a single real scalar field $\Phi$ in curved space,
\beq
S = \int d^4x \sqrt{-g}\,\Biggl\{\sum_{k=1}^{N}
\bar{\Psi}_k(i\gamma^{\mu}\na_\mu-h\Phi)\Psi_k
+ \frac{1}{2}(\na_{\mu} \Phi)^{2}
+ \frac{1}{2}\xi R\Phi^{2}
- \frac{\la}{4!}\Phi^{4}
+ \tau \cx \Phi^2
\Biggr\},
\label{S0}
\eeq
with nonminimal parameter $ \xi=\frac16$. Other couplings include
scalar self-interaction $\la $ and Yukawa constant $h$. The conformal
model (\ref{S0}) is useful owing to its simplicity and generality. The
last means the presence of fermions with Yukawa interaction provide
renormalization of the kinetic term for the scalar field already at
the one-loop level. The total derivative term $\cx \Phi^2$ with
arbitrary parameter $\tau$ is required to have renormalizable
theory. The renormalization of this term is very important, as we
shall see in what follows.

The generalization to an arbitrary conformal multiscalar model
with spinor, vector fields, and an arbitrary gauge group is not
complicated. The general action has the form $S(\Phi,\Psi,A,g)$,
where $\Phi$, $\Psi$, $A$ and $g$ are scalars, fermions, vectors
and metric.

On top of the general covariance, this model is also invariant under
the transformation called local conformal symmetry,
\beq
g_{\mu\nu}=e^{2\si}\,{\bar g}_{\mu\nu},
\quad
\Phi=e^{-\si}\,{\bar \Phi},
\quad
\Psi=e^{-\frac32\si}\,\Psi_{*},
\quad
\bar{\Psi}=e^{-\frac32\si}\,\bar{\Psi}_{*},
\quad
A=\bar{A},
\quad
\si=\si(x).
\label{conf-4}
\eeq

The renormalizable theory on curved backgrounds (see, e.g.
\cite{book,OUP} for the introduction) requires the conformal
vacuum term
\beq
S_{cv}=\int d^{4}x \sqrt{-g}\;
\Big\{a_1C^{2}+a_2E_4+a_3\cx R\Big\},
\label{Svac}
\eeq
where
\beq
C^2(4)\,=\,R_{\al\be\mu\nu}R^{\al\be\mu\nu}
- 2 R_{\al\be}R^{\al\be} + \frac13 R^2
\label{Weyl}
\eeq
is the square of the Weyl tensor and
$E_4\,=\,R_{\al\be\mu\nu}R^{\al\be\mu\nu}
- 4 R_{\al\be}R^{\al\be} + R^2$
is the integrand of the Gauss-Bonnet topological term.

Regardless the surface terms $\cx \Phi^2$, $E_4$, and $\cx R$ are
not conformal, the symmetry holds in the corresponding Noether
identity corresponding to (\ref{conf-4}),
\beq
\mathcal{T}
\,=\,
-\,\frac{2}{\sqrt{-g}}\;g_{\mu\nu}\,
\frac{\de S(g_{\mu\nu},\Phi)}{\de g_{\mu\nu}}
\,-\,\frac{d_\Phi}{\sqrt{-g}}\,\Phi
\,\frac{\de S(g_{\mu\nu},\Phi)}{\de \Phi}
\,=\, 0
%%  red
%%\textcolor{red}{
,
\label{Noether_ident}
\eeq
where $d_\Phi=-1$ is the conformal weight of the background scalar field.
If including fermions and vectors, the weights are  $d_\Psi=-\frac32$ and
 $d_A=0$.

According to the existing general proof \cite{tmf84} (see also
\cite{OUP} for a simplified version), the one-loop divergence
$\bar{\Ga}^{(1)}_{div}$ in the theory (\ref{S0}) is conformal
invariant. This means, in dimensional regularization,
\beq
\bar{\Gamma}^{(1)}_{div}
\,=\,-\,\dfrac{1}{\varepsilon}\int d^{4}x \sqrt{-g}\;\Delta \mathcal{L}_c,
\label{ctr}
\eeq
where $\varepsilon \equiv (4\pi)^{2}(n-4)$ and the local functional
${\displaystyle\int} d^{4}x \sqrt{-g}\;\Delta \mathcal{L}_c $
is conformal i.e., satisfies \eqref{Noether_ident}.

On the other hand, the renormalized one-loop effective action
\beq
\Gamma^{(1)}_{\textrm{ren}}
\,=\,
S+S_{cv}+\bar{\Gamma}^{(1)}+\Delta S^{(1)},
\label{G_full}
\eeq
violates Noether identity. Here
$\bar{\Ga}^{(1)}=\bar{\Ga}^{(1)}_{div}+\bar{\Ga}^{(1)}_{fin} $ is
non-renormalized one-loop contribution and $\Delta S^{(1)}$ includes
local counterterms introduced to cancel the UV divergences. The
breaking of the symmetry due to quantum corrections characterizes
the conformal anomaly \cite{CapDuf-74,ddi,duff77}.

%%%%%%%%%%%%%%%%%%%%%%%%%%%%%%%%%%%%
The expression for the divergences in our model (\ref{S0}) is
\beq
&&
\bar{\Gamma}^{(1)}_{div} \,=\, -\, \frac{\mu^{n-4}}{\vp}
\int d^{n}x\,\sqrt{-g}\,\biggl\{\frac{1+6N}{120}\,C^{2}
- \frac{1+11N}{360}E_4+2Nh^{2}\Big[(\na \Phi)^2
+ \frac16\,R\Phi^{2}\Big]
\nn
\\
&&
\qquad\quad
+\,\,\frac{1+6N}{180}\,\square R+\Big(\frac{\la^2}{8}-2Nh^4\Big)\Phi^4
+\Big(\frac{\la}{12}-\frac{4Nh^{2}}{3}\Big)\cx \Phi^{2}\biggr\},
\label{div_1loop}
\eeq
where $(\na\Phi)^2 = g^{\mu\nu}\pa_\mu\Phi\,\pa_\nu\Phi$.
In the limit $n \to 4$, the integral in the expression (\ref{div_1loop})
satisfies \eqref{Noether_ident}.

In the general case, when the theory includes scalars, spinors
and vectors, the one-loop conformal invariance implies that the
one-loop divergences have the form
\beq
&&
\bar{\Gamma}^{(1)}_{div} = - \frac{\mu^{n-4}}{n-4}
\int d^{n}x\,\sqrt{-g}\,\biggl\{
wC^2 + bE_4 + c \cx R
\nn
\\
&&
\qquad
\,\,\,
-\,
%% \frac{1}{2}
\,\gamma_\Phi\Big[(\nabla\Phi)^2 +\frac16R\Phi^2 \Big]
\,+\,\frac{1}{4!}\,\widetilde{\be}_{\la} \Phi^4
\,\,+\,\be_\tau \square\Phi^2\biggr\},
\label{Gamma1loop}
\eeq
where $\widetilde{\be}_{\la} =\be_\la + 4\la\gamma_\Phi$.  The gamma
function $\gamma_\Phi$ and the beta functions $w$, $b$, $c$,
$\be_\la$ and $\be_\tau$ depend on the particle contents of the theory.

The coefficients $w$, $b$ and $c$ are the one-loop $\be$-functions
in the vacuum sector. They depend only on the number of fields of
different spins \cite{birdav,OUP},
\beq
&&
w = \frac{1}{(4\pi)^2}\,\bigg(
\frac{N_0}{120} + \frac{N_{1/2}}{20} + \frac{N_1}{10} \bigg),
\label{w}
\\
&&
b = -\,\frac{1}{(4\pi)^2}\,\bigg( \frac{N_0}{360}
+ \frac{11\,N_{1/2}}{360} + \frac{31\,N_1}{180}\bigg),
\label{b}
\\
&&
c = \frac{1}{(4\pi)^2}\,\bigg( \frac{N_0}{180} + \frac{N_{1/2}}{30}
- \frac{N_1}{10}\bigg).
\label{c}
\eeq
In the particular case of \eqref{div_1loop}, $N_0=1$, $N_{1/2}=N$
and $N_1=0$. In the scalar sector, $\be_\la$, $\gamma_\Phi$ and $\be_\tau$
depend on the gauge group, coupling constants and other details of
the model. In the simple example (\ref{div_1loop}) considered above,
\beq
&&
\gamma_\Phi = -\dfrac{1}{(4\pi)^{2}}\,2N h^{2},
\label{gamma}
\\
&&
\widetilde{\be}_{\la}=\dfrac{1}{(4\pi)^{2}}\big(3\la^{2}-48h^4N\big),
\label{la}
\\
&&
\be_{\ta}=\dfrac{1}{(4\pi)^{2}}\bigg(\dfrac{\la}{12}
-\dfrac{4N h^2}{3}\bigg).
\label{4}
\eeq

The anomaly cannot be derived completely using the identity
\eqref{Noether_ident} because of the problems with surface
terms \cite{birdav,duff94}. However, following the scheme of
\cite{anomaly-2004,PoImpo}, one can obtain the anomaly by
taking variational derivative with respect to the conformal factor,
\beq
&&
\langle \mathcal{T} \rangle
\,=\, - \frac{1}{\sqrt{-\bar{g}}}\,e^{-4\si}\,
\frac{\de {\bar\Ga}^{(1)}_{\textrm{ren}}}{\de \si}\bigg|
\,=\,
 - \,\frac{1}{\sqrt{-\bar{g}}}\,e^{-4\si}\,
\frac{\de \De S^{(1)}}{\de \si}\bigg| \nn \\
&&
\qquad
= \,\,-\, wC^2 - bE_4 - c \cx R
\,-\, \frac{1}{4!}\,\widetilde{\be}_\la \Phi^4
\,-\, \be_\tau\cx\Phi^2
\,+\, \gamma_\Phi
\Big[(\nabla\Phi)^2 +\frac16R\Phi^2 \Big].
\label{anomaly}
\eeq
Here $ \big| $ means the replacement   %%   procedure
$\bar{g}_{\mu\nu} \rightarrow g_{\mu\nu}$,
$\bar{\Phi} \rightarrow \Phi$, and $\si \rightarrow 0$.

It is remarkable that the general form of the divergences in the
scalar sector remains qualitatively the same (\ref{div_1loop}) in
any other conformal model with a scalar field or fields. Thus, the
general structure of the anomaly \eqref{anomaly} holds on. The
main modification concerns corresponding renormalization group
functions. In particular, there are no dramatic changes if the scalar
is complex and multi-component. For the MSM Higgs we have to
make the following replacements in both divergences and anomaly:
\beq
&&
\frac{1}{2}(\pa \Phi)^{2} \,\rightarrow\, g^{\mu\nu}
{\mathcal D}_\mu H^{\dagger} {\mathcal D}_\nu H,
\qquad
\frac{1}{12}R\Phi^{2} \,\rightarrow\, \frac{1}{6}RH^{\dagger}H,
\nn
\\
&&
\frac{1}{4!}\,\Phi^{4}
\,\rightarrow\, \frac{1}{8}(H^{\dagger}H)^{2},
\qquad
\qquad
\quad
\cx \Phi^{2} \,\rightarrow\, \cx (H^{\dagger}H)
\label{SMS-case}
\eeq
and the rest is the same.

As another example, consider the $SU(2)$ model \cite{VoTyu76}
\beq
&&
S \,=\, \int d^{4}x\,\sqrt{-g}\,\biggl\{
- \frac{1}{4}G^{a}_{\mu\nu}G^{a\mu\nu}
\,+\,
%%  red
%%\textcolor{red}{
i
%% \sum_{k=1}^{s}
{\bar\Psi}^{a}%% _{k}
\Big(\ga^{\mu} {\mathcal D}^{ab}_{\mu} - h\varepsilon^{acb}\Phi^c
\Big)\Psi^{b}%% _{k}
\nn
\\
&&
\qquad
+\,\,  \frac{1}{2}\,
g^{\mu\nu}({\mathcal D}_\mu\Phi)^{a}({\mathcal D}_\nu\Phi)^{a}
+ \frac{1}{12}\, R\Phi^{a}\Phi^{a}-\frac{1}{4!}\la(\Phi^a\Phi^a)^2+\ta\cx(\Phi^a\Phi^a)
\biggr\},
\label{mod_SU2}
\eeq
%%  red
%%\textcolor{red}{
where scalars and spinors are in the adjoint representation
of the gauge group, %% $s$ is the number of spinor multiplets,
$ G^{a}_{\mu\nu}=\nabla_\mu A^{a}_{\nu}-\nabla_\nu A^{a}_{\mu}+g\varepsilon^{abc}A^{b}_{\mu}A^{c}_{\nu}$,
and  $({\mathcal D}_{\mu}\Phi)^{a}=\de^{ab}\na_\mu\Phi^{b}
+ig\vp^{acb}A^{c}_{\mu}\Phi^{b} $
is the covariant derivative of scalars (or the same for spinors).
In this theory, the coefficients in the scalar sector are \cite{book}
\beq
&&
\gamma_\Phi = - \dfrac{4}{(4\pi)^{2}}(h^{2}-g^{2}),
\label{gamma2}
\\
&&
\widetilde{\be}_{\la} = \dfrac{1}{(4\pi)^{2}}
\bigg(\dfrac{11}{3}\la^{2}-8\lambda g^{2}+72g^{4}-96h^{4}\bigg),
\label{la2}
\\
&&
\be_\tau = \dfrac{1}{(4\pi)^{2}}\bigg(\dfrac{5}{36}\la
+ \dfrac{2}{3}g^{2}-
%%  red
%%\textcolor{red}{
\dfrac83
h^{2}\bigg).
\label{42}
\eeq
One can also find the coefficients for other cases, such as
GUT-like models (e.g. \cite{GUT-90}) and the Minimal Standard
Model \cite{2Yoon}. As we have mentioned above, the general
structure of anomaly (\ref{anomaly}) remains and only the
coefficients get modified.

At this point we can make a classification of the scalar-dependent
structures in the anomaly (\ref{anomaly}), similar to that of
\cite{ddi} and \cite{DeserSchwimmer} in the purely metric case.
There are

$i)$ Real conformal terms, such as $C^2$, $\Phi^4$ and
$\,(\na \Phi)^2 + \frac{1}{6}R\Phi^2$. It proves useful
introducing special notation for the generalized conformal
structures in the anomaly,
\beq
&&
X_c \,=\,
%% \frac12 \Big[
(\na\Phi)^2  +  \dfrac{1}{6}\,R\Phi^2
%%  red
%%\textcolor{red}{
, %% \Big],
\nn
\\
&&
Y(g_{\mu\nu},\Phi)
\,=\,
wC^2 - %% \dfrac{1}{2}\,
\gamma_\Phi X_c
%% \Big[ (\na\Phi)^2  +  \dfrac{1}{6}\,R\Phi^{2}\Big]
 + \dfrac{1}{4!}\,\widetilde{\be}_{\la}\Phi^4.
\label{Y}
\eeq

$ii)$ Unique topological term $E_4$, independent on the extra
fields such as scalars.

$iii)$ Total derivatives, in the present case  $\Box R$ and
$\Box\Phi^2$.  In the next section we shall see that these terms are
generated by local terms in the induced action. Whether or not these
terms can be regarded as irrelevant, depends on the model, as we
shall discuss in what follows.

Let me mention that the topological term remains unique in
higher (even, at least) dimensions \cite{6d}. However, there
is no general proof that the total derivatives in the anomaly can
be always generated by local actions regardless this is the case
in dimension six and, in general, for all available examples.

%%%%%%%%%%%%%%%%%%%%%%%%%%%%%%%%
%%%%%%%%%%%%%%%%%%%%%%%%%%%%%%%%
%%%%%%%%%%%%%%%%%%%%%%%%%%%%%%%%
\section{Integration of anomaly with a scalar field}
\label{sec4}

To anomaly-induced effective action of the background
fields $g_{\mu\nu}$ and $\Phi$ is a solution to the equation
\beq
- \frac{2}{\sqrt{-g}}\,g_{\mu\nu}\,
\frac{\de \Ga_{ind}}{\de g_{\mu\nu}}
\,+\, \frac{1}{\sqrt{-g}}\,\Phi \,\frac{\de \Ga_{ind}}{\de \Phi}
\,=\,
- \frac{1}{\sqrt{-\bar{g}}}\,e^{-4\si}\,
\frac{\de \Ga_{ind}}{\de \si}\bigg|
\,\,=\,\, \langle \mathcal{T} \rangle
%%  red
%%\textcolor{red}{
.
\label{EA}
\eeq

Such a solution for a purely gravitational case was found in
\cite{rie,frts84} and represents a four dimensional generalization of
the Polyakov action \cite{polyakov81} in two dimensions. There are
generalizations including torsion \cite{BuchOdShp85,anhesh} and
the parity-violating term \cite{MaurShp15}. Also, the general
solution for an arbitrary even dimension was obtained in \cite{6d}.

The first observation is that the total derivative terms in the anomaly
$\langle \mathcal{T} \rangle $ can be obtained using the relations
\beq
&&
-\frac{2}{\sqrt{-g}}g_{\mu\nu}\frac{\delta}{\delta g_{\mu\nu}}
\int d^4 x\sqrt{-g}\,R^{2}\,=\,12\square R
%%  red
%%\textcolor{red}{
,
\label{TrR2}
\\
&&
\biggl(-\frac{2}{\sqrt{-g}}g_{\mu\nu}\frac{\delta}{\delta g_{\mu\nu}}
+\frac{1}{\sqrt{-g}}\Phi\frac{\delta}{\delta \Phi}\biggr)
\int d^4 x\sqrt{-g}\,R\Phi^{2}\,=\,6\square \Phi^{2}.
\label{RPhi2}
\eeq

The conformal invariants %%  such as $\Phi^4$ or $X_c$,
can be kept together with $C^2$ as in Eq.~(\ref{Y}), simplifying a lot
the derivation of anomaly-induced action. Furthermore, we will need
the relation
\beq
\sqrt{-g}\,\Big(E_4-\frac23\,{\cx} R\Big)
\,=\,
\sqrt{-\bar{g}}\,\Big({\bar E_4}-\frac23\,{\bar \cx} {\bar R}
+ 4{\bar \De_4}\si \Big),
\label{119}
\eeq
where $ \Delta_4 $ is the Paneitz operator
\cite{FrTs-superconf,Paneitz}
\beq
\Delta_4\,=\,
\cx^2 + 2R^{\mu\nu}\nabla_{\mu}\nabla_{\nu} - \dfrac{2}{3}R\square
+\dfrac{1}{3}(\nabla^{\mu}R)\nabla_{\mu},
\label{Paneitz_op}
\eeq
satisfying $\sqrt{-g}\Delta_4=\sqrt{-\bar{g}}\bar{\Delta}_4$.
In what follows we use the compact notation
$\int_x\equiv \int d^4 x\sqrt{-g(x)}$.
The fundamental relation (\ref{119}) is the main element that enables
the integration of anomaly, as it provides the formula
\beq
\frac{\de}{\de\si}
\int_xA\Big(E_4-\frac23 \square R\Big)\bigg|=4\sqrt{-g}\De_4 A,
\label{Conf_functional}
\eeq
that is valid for an arbitrary conformal functional
$A[g_{\mu\nu},\Phi]=A[\bar{g}_{\mu\nu},\bar{\Phi}]$.

The simplest solution can be obtained directly from (\ref{EA}),
(\ref{TrR2}), (\ref{RPhi2}) and  (\ref{119}), in the form
\beq
&&
\Ga_{ind}\,=\,
S_c[{\bar g}_{\mu\nu},{\bar \Phi}]
\,-\,\int d^4 x\sqrt{-g}\,\bigg\{\frac{2 b+3c}{36}R^2
+ \frac{\be_\tau}{6}R\Phi^2\bigg\}
\nn
\\
&&
\qquad
\quad
\,+\,\int d^4 x \sqrt{-{\bar g}}
\,\bigg\{\si  Y\big(\bar{g}_{\mu\nu},\bar{\Phi}\big)
\,+\, b\si \Big({\bar E}-\frac23 {\bar \cx}
{\bar R}\Big) + 2b\si{\bar \De}_4\si\bigg\},
\label{quantum1}
\eeq
where $\sqrt{-g} = \sqrt{-{\bar g}} e^{4\si}$ and
$R = e^{-2\si}\big[{\bar R}
- 6({\bar \na}\si)^2 - 6 \bar {\cx} \si \big]$. In (\ref{quantum1}),
$S_c[{\bar g}_{\mu\nu},{\bar \Phi}]=S_c[g_{\mu\nu},\Phi]$
is an ``integration constant'' for equation \eqref{quantum1} i.e.,
the conformal functional that cannot be obtained from anomaly
and, therefore, has no direct relation to the UV divergences.

An alternative, non-local covariant solution of \eqref{EA}
requires introducing Green function for the Paneitz operator,
\beq
(\sqrt{-g}\De_4)_xG(x,y)=\de(x,y).
\label{Green_function}
\eeq
Using
(\ref{Conf_functional}) and the general scheme \cite{6d},
we obtain (see e.g. \cite{OUP} for the details)
\beq
&&
\Ga_{ind}
\,=\,
S_c
\,+\,
\frac{b}{8}\int_x\int_y\Big(E_4
-\frac23\square R\Big)_{\hspace{-1mm}x}
G(x,y)\Big(E_4-\frac23\square R\Big)_{\hspace{-1mm}y}
\nn
\\
&&
\qquad \quad
+\,
\frac{1}{4}\int_x\int_y Y(x)\, G(x,y)
\Big(E_4-\frac23\square R\Big)_{\hspace{-1mm}y}
\,-\,
\int_x\,\Big(\frac{2 b+3c}{36}R^2
+ \frac{\be_\tau}{6} R\Phi^2\Big).
\label{nonlocal}
\eeq

Rewriting (\ref{nonlocal}) in the symmetric form, we get
\beq
&&
\Gamma_{ind}
\,=\,
S_c
\,-\,\int_x\,\Big(\frac{2 b+3c}{36}R^2
+ \frac{\be_\tau}{6} R\Phi^2\Big)
- \frac{1}{8b}\int_x\int_y\;Y(x)G(x,y)Y(y)
\nn
\\
&&
\qquad
\quad
+\,\frac{b}{8}\int_x\int_y\Big(E_4-\frac23\square R+\frac{1}{b}Y\Big)_{\hspace{-1mm}x}G(x,y)\Big(E_4-\frac23\square R+\frac{1}{b}Y\Big)_{\hspace{-1mm}y}.
\label{nonlocal2}
\eeq
Finally, we rewrite the result \eqref{nonlocal2} in the local
representation by introducing two auxiliary fields $\varphi$ and
$\psi$ \cite{a} (see also \cite{MazMott01,MaurShp15} for an
alternative presentation). The result is
\beq
&&
\Gamma_{ind}
\,=\,
S_c[g_{\mu\nu},\Phi]
\,- \int_{x}\bigg\{\frac{2 b+3c}{36}R^2
+ \frac{\be_\tau}{6} R\Phi^{2}\bigg\}
+\int_{x}\bigg\{\frac12 \varphi\Delta_4\varphi-\frac12 \psi\Delta_4\psi
\nn
\\
&&
\qquad\quad
+\,\,
\frac{\sqrt{-b}}{2}\varphi
\Big(E_4-\frac23\square R+\frac{1}{b}Y\Big)
\,+\, \frac{1}{2\sqrt{-b}}\,\psi Y
\bigg\}.
\label{aux_fields}
\eeq
%% red
%%\textcolor{red}{
Let us note that the non-local covariant
expression (\ref{nonlocal}) is a particular case of the
formula given in the recent work \cite{Mottola-2017}
(there were also particular examples for other fields, e.g.,
\cite{GianMott09,MaurShp15}).
As we already mentioned above, this structure is quite general
and actually holds in any even-dimensional spacetime \cite{6d}.
%%  end of red
%%  end of red
%%  end of red
A remarkable feature of the solutions (\ref{nonlocal})
and (\ref{aux_fields}) is that the induced effective action,
as given in these formulas, is invariant under global conformal
transformations (\ref{conf-4}),  i.e., with $\si=const$.
However, this invariance does not contradict the fact that
the global conformal symmetry of the theory is anomalous.

The forms  \eqref{quantum1}, \eqref{nonlocal} and \eqref{aux_fields}
are equivalent, regardless for the non-cosmological applications (see,
e.g., \cite{PoImpo} and references therein) \eqref{aux_fields} is more
useful owing to covariance and locality. The phenomenological
generalization for light massive quantum fields has been constructed
in \cite{shocom,asta} for the simplest form \eqref{quantum1}.

Concerning zero-order cosmology i.e., for the dynamics of the
conformal factor of the metric, we note that there is a significant
difference between the purely metric background and the solution
with scalar, such as (\ref{aux_fields}). In the metric case, the
functional $S_c$ is irrelevant because the metric is defined by
(\ref{conf-4}) with $\,e^\si = a(\eta)$, where
$\eta$ is conformal time. As a result, $S_c$ does not depend on
$a(\eta)$, and the anomaly-induced effective action looks like
an exact form of quantum corrections at the one-loop level.
Furthermore, assuming that the general structure of anomaly does
not change at higher loops, the unique detail that may change
beyond the one-loop order concerns the beta functions. In the scalar
case, anomaly includes extra terms (\ref{anomaly}) and, on the
other hand, the functional $S_c$ depends on the scalar fields
$\Phi$. Thus, the solution (\ref{aux_fields}) cannot be regarded as
exact anymore. The status of this solution is like one of the purely
metric background in the case of black hole solution or other
similar situations \cite{PoImpo}.

Another important question is whether the form of anomaly may
be the same at higher loops, if the initial quantum theory includes
scalar fields. To address this issue we have to explore in detail
the effective action of scalar field at one loop, including the
corresponding ambiguities. This will be the subject of the next
section.

The last important item to note is that the expressions
\eqref{quantum1}, \eqref{nonlocal} and \eqref{aux_fields} are valid
only for massless conformal fields. In case of massive Higgs or other
massive fields contributing to $\widetilde{\be}_\la$, $\gamma_\Phi$,
$w$, $b$, $c$ and $\be_\tau$, these results can be regarded as UV
approximation. In the case of $\widetilde{\be}_\la$, $\gamma_\Phi$,
$w$ and $c$ this fact can be confirmed by direct analysis of the
nonlocal form factors \cite{apco,bexi}.

%%%%%%%%%%%%%%%%%%%%%%%%%%%%%%%%
%%%%%%%%%%%%%%%%%%%%%%%%%%%%%%%%
%%%%%%%%%%%%%%%%%%%%%%%%%%%%%%%%
\section{Ambiguities in the total derivative terms}
\label{sec3}

Let us discuss the existence of ambiguities in the beta functions of
the surface terms $c$ and $\be_\tau$ of \eqref{anomaly}. We already
know from (\ref{TrR2}) and (\ref{RPhi2}) that these terms produce
$\int d^4x\sqrt{-g}R^2$ and $\int d^4x\sqrt{-g}R\Phi^2$ finite terms
in the induced action. %%  and thus can be quite relevant.

The case of the $c\cx R$-term was extensively discussed in the
literature as mainly unsolved problem coming from the choice of
regularization. In this respect, it is worth mentioning the book
\cite{birdav} and important review paper \cite{duff94}. In a more
conclusive way, the problem was considered in
\cite{anomaly-2004} for the particular example of a real scalar
field. The ambiguity in the coefficient $\be_\tau$ has been
considered in \cite{BoxAno} for the two examples, i.e., the
self-interacting scalar and the Yukawa model. In the last case the
theory had a purely background scalar field and took into account
quantum effects of a fermion. The analysis in
Ref.~\cite{anomaly-2004} included various types of regularization.
One of the main points was that there is an ambiguity in the choice
of the Weyl-squared counterterm, related to the possibility to choose
the counterterm with $C^2(d)$  with $d=n+\gamma(n-4)$, instead of
$C^2(4)$ \cite{duff77} or a more simple version with $C^2(n)$.
It was shown that this ambiguity is completely equivalent to the
freedom of introducing a \textit{finite} $\int d^4x\sqrt{-g}R^2$-term
in the \textit{classical} vacuum action (\ref{Svac}). As this term
produces $\cx R$ in the Noether identity (\ref{Noether_ident}),
the $\cx R$-term in the anomaly gets modified and looks ambiguous.
Another example of a qualitatively similar ambiguity, not related to
the choice of $C^2(d)$, is the Pauli-Villars regularization, that
shows an ambiguity for the beta function $c$ \cite{BoxAno}.

The important difference between $c$ and $\be_\tau$ is that
introducing an $R^2$-term into the lagrangian of vacuum does not
spoil the symmetry for the quantum fields and can be regarded as
kind of ``legitimate'' procedure. The situation is opposite for the
$\Phi^2$-term. Changing the coefficient of this term in the classical
lagrangian from $\frac16$ to any $\xi\neq \frac16$ breaks down the
conformal symmetry in the sector of quantum fields. Then, the
one-loop divergences become conformally non-invariant i.e., they do
not have special structure of (\ref{Gamma1loop}). In this case, one
needs an independent renormalization of $\xi$ in the action
(\ref{S0}), hence the conformal value $\xi = \frac16$ cannot be a
fixed point of the renormalization group flow. Strictly speaking,
without conformal symmetry there is no sense to speak about anomaly.
Thus, the $\cx \Phi^2$-type ambiguity is a critical issue for the
quantization of any quantum theory with scalar fields.

%%  As we already mentioned,
Since for the $\cx \Phi^2$ term %% in the anomaly
there is no ambiguity in the dimensional regularization,
%% . To explore the ambiguity in other regularizations,
in what follows we extend the Pauli-Villars - based analysis
\cite{anomaly-2004,BoxAno}
to arbitrary model with scalars, fermions and vectors. As we shall
see below, this extension is not completely trivial. As before, we
assume that there is only a single real scalar $\Phi$, but without
imposing restrictions on gamma and beta functions in the
divergences (\ref{Gamma1loop}). %%  anomaly (\ref{anomaly}).

Pauli-Villars regularization requires introducing the set of scalar
or spinor fields (regulators) \cite{Slavnov-PV,AsoFalc} with the
specially chosen masses $m_i$ and with indefinite Grassmann
parity. We will need the one-loop contributions of an individual
scalar regulator $\varphi_i$ with the mass $m_i$ in the external
background of the scalar field $\Phi$ and metric,
\beq
S^{(i)}_{reg} &=&
\int d^4 x\sqrt{-g}
\,\Big\{\,\frac12\,g^{\mu\nu}
\partial_\mu\,
\varphi_i\partial_\nu\varphi_i +\frac{\xi_i}{2}\,R\,\varphi_i^2
\,-\,\frac{m^2_i}{2}\, \varphi_i^2\, -\frac{\ka}{2}
\Phi^2\varphi_i^2 \Big\},
\label{regscal}
\eeq
where $\ka$ is an artificial coupling which will prove useful later
on, while $m_i$ and $\xi_i$ are masses and non-minimal parameters
for the scalars regulators.

The contribution of the regulator may enter with either positive or
negative sign, depending on the statistics of $\varphi_i$. In both
cases, we can use the expression for the effective action which was
calculated using the heat kernel method \cite{bavi90,Avra89} in
\cite{apco,Omar4D}, including for the metric-scalar
background \cite{bexi}
%% red
%%\textcolor{red}{
(an independent equivalent calculation was reported in
\cite{Meissner-2008})
%%%%%%%%%%%%%%%  end red
%%%%%%%%%%%%%%%  end red
%%%%%%%%%%%%%%%  end red
and also using Feynman diagrams
\cite{apco,CodelloZanusso2013,OUP}. The result, for the bosonic
statistics, has the form
%%
%% DO SIH
\beq
&&
\bar{\Ga}^{(1)}_{scal}
\,=\, \frac{1}{2(4\pi)^{2}}
\int d^4x\,\sqrt{-g}\,
\biggl\{m_i^4\Big(\frac{1}{2\ep} + \frac34\Big)
+\widetilde{\xi}_i\, m_i^2 R \Big(\frac{1}{\ep}+1\Big)
\nn
\\
&&
\qquad \quad
+ \,\,C_{\mu\nu\alpha\beta}\Big[\frac{1}{120\ep}
+ \frac12\,k_W(\tau_i)\Big]C^{\mu\nu\alpha\beta}
+R\Big[\frac{1}{2\ep}\,\widetilde{\xi}_i^{\,2}+k_R(\tau_i)\Big]R
\nn
\\
&& \qquad \quad
- \,\,
\frac{\ka}{2\ep}\,m_i^2\Phi^2
+ \Phi^{2}\Big[\frac{\ka^2}{8\ep}
+ k_\ka(\tau_i)\Big]\Phi^2
+ \Phi^2
\Big[-\frac{\ka}{2\ep}\,\widetilde{\xi_i} + k_\xi(\tau_i)\Big]R \biggr\},
\label{Gsc}
\eeq
where $\tau_i = \cx/m_i^2$ and we use the compact notations
\beq
\widetilde{\xi}_i=\Big(\xi_i-\frac16\Big),
\qquad
\frac1\ep \equiv \frac{2}{4-n}
+\ln\Big(\frac{4\pi\mu^2}{m^2}\Big)-\ga,
\eeq
where $\ga \approx 0.577$ is the Euler-Mascheroni constant. The
expressions for the non-local form factors $k_W$,  $k_R$, $k_\ka$
and $k_\xi$ can be found in \cite{bexi}. For us it is sufficient to
remember that (i) the UV logarithmic factors are always proportional
to $\ln \mu^2$, as shown below; \ \ (ii) in the IR limit $m_i \to \infty$
all these form factors vanish as $\,{\mathcal O}(\tau_i)$.

In the conformal limit, $m_i \rightarrow 0$ and
$\widetilde{\xi}_i \to 0$,  the finite part in \eqref{Gsc}
boils down to
\beq
&&
\bar{\Ga}^{(1)}_{UV}
\,\,=\,\,
-\frac{1}{2(4\pi)^{2}}\int d^{4}x\,\sqrt{-g}\,
\biggl\{
\frac{1}{120}\,
C_{\mu\nu\alpha\beta}\ln\Big(- \frac{\cx}{4\pi\mu^{2}}\Big)
C^{\mu\nu\alpha\beta}
\nn
\\
&&
\qquad\quad
\,+\,\,\frac{\ka^2}{8}\,
\Phi^{2}\ln\Big(- \frac{\cx}{4\pi\mu^{2}}\Big)\Phi^{2}
\,+ \,\frac{1}{1080}\,R^2
\,+\, \frac{\ka}{36}\,\Phi^2 R\biggr\}.
\label{G_conf}
\eeq
All the terms here violate local conformal symmetry. The first
two terms in the integrand are nonlocal and correspond to the one-loop
divergences in the Weyl-squared and $\Phi^4$ terms, respectively.
The last two terms are local and correspond to the total derivative,
$\cx R$ and  $\cx \Phi^2$ terms in the anomaly via (\ref{TrR2})
and (\ref{RPhi2}). One can say that (\ref{G_conf}) is an alternative
to (\ref{nonlocal2}) form of the effective action responsible for
anomaly \cite{BMZ}, but we shall use this expression as an
instrument to explore the ambiguity.

The procedure of Pauli-Villars regularization starts from
(\ref{Gsc}) and requires the set of auxiliary (regulator) fields
(\ref{regscal}) with $i = 1,...,N$. For exploring the ambiguity,
these fields should have general $\widetilde{\xi}_i$. Each field
has a degeneracy $s_i$ multiplied by $1$ in the case of bosonic,
or $-2$ for the fermionic statistics. Starting from the simplest case
of a single scalar $\Phi$, we label it as $\ph_0$, assuming
$\widetilde{\xi}_0=0$ and $s_0=1$.
%% DO SIH
%% \vskip 6mm
Then the Pauli-Villars regularized effective action can be defined as
\beq
&&
{\bar\Ga}^{(1)}_{\textrm{reg}} = %% \lim_{\La \rightarrow \infty}
\sum^{N}_{i=0}s_i{\bar\Ga}^{(1)}_{i}(m_i,\,\widetilde{\xi}_i,\,n).
\label{EA_reg}
\eeq
%% where $\La$ is an auxiliary cut-off.
The $i=0$ term is given by \eqref{G_conf} plus the divergent part
${\mathcal O}(1/\ep)$. Consider $m_i=\mu_i M$,
where $M$ is the dimensional parameter of regularization and
$\mu_i$ are dimensionless coefficients to be defined. We can choose
$\mu_i$ in such a way that the ${\mathcal O}(1/\ep)$
terms in (\ref{EA_reg}) cancel out and the auxiliary dimensional
regularization becomes irrelevant. In fact, we could replace it by the
covariant cut-off in the heat-kernel integral, as it was done in
\cite{anomaly-2004,BoxAno}.

The Pauli-Villars conditions eliminating ${\mathcal O}(1/\ep)$ terms
and (in case of covariant cut-off) also the quadratic and quartic
divergences, have the form
\beq
&&
\sum^{N}_{i=1}s_i=-s_0=-1;
\label{PVcond1}
\\
&&
\sum^{N}_{i=1}s_i\mu^{2}_i=0,
\;\;\;\;\; \sum^{N}_{i=1}s_i\,{\widetilde\xi}_i=0;
\label{PVcond2}
\\
&&
\sum^{N}_{i=1}s_i\mu^{4}_i=0,
\;\;\;\;\; \sum^{N}_{i=1}s_i\,{\widetilde\xi}^{\,2}_i=0.
\label{PVcond3}
\eeq
Some explanation is in order. The condition \eqref{PVcond1}
provides cancellation of quartic divergences and also Weyl-squared
${\mathcal O}(1/\ep)$ terms. The two conditions \eqref{PVcond2}
eliminate quadratic divergences in the cut-off scheme. Finally,
the conditions \eqref{PVcond3} eliminate logarithmic divergences
(i.e., ${\mathcal O}(1/\ep)$) of the cosmological constant and
$R^2$-types.

A possible solution to these conditions corresponds to $N=5$ and
\beq
&&
s_1=1, \qquad s_2=4, \qquad s_3=s_4=s_5=-2;
\label{sol1}
\\
&&
\mu^2_1=\mu^2_5=4,
\qquad
\mu^2_2=\mu^2_4=3,
\qquad
\mu^2_3=1;
\label{sol2}
\\
&&
\widetilde{\xi}_i=\mu^2_i.
\label{sol3}
\eeq
This solution is also valid for combinations of the conditions in
\eqref{PVcond2} and \eqref{PVcond3}, i.e.,
\beq
&&
\sum^{N}_{i=1}s_i\mu^2_i\,{\widetilde\xi}_i=0,
%% \;\; \textrm{and} \;\; \sum^{N}_{i=1}s_i\mu^{4}_i\,{\widetilde\xi}^{\,2}_i=0.
\label{PVcond}
\eeq
required for the $R\Phi^2$ divergence. Using \eqref{EA_reg}, we
arrive at the following expression for conformal anomaly in the
covariant Pauli-Villars regularization for a single scalar field,
\beq
&&
\langle \mathcal{T} \rangle
\,=\,
-\frac{\be_\la}{4!}\Phi^4
+ %% \frac12
\gamma_\Phi X_c
%%  \frac12 \Big[\big(\na \Phi\big)^2 + \frac16\,R\Phi^2\Big]
- wC^2 - bE_4
- (c - 6\delta) \square R
- (\be_\tau + 3\rho) \square\Phi^2,
\,\,\qquad
\label{anomaly_amb}
\eeq
where $\ga_\Phi = 0$ and we define
\beq
&&
\rho \,=\, \frac{1}{2(4\pi)^2}
\sum^{N}_{i=1}s_i\,{\widetilde\xi}_i\ln \mu^2_i;
\label{PVcoef1}
\\
&&
\delta \,=\, \frac{1}{2(4\pi)^2}
\sum^{N}_{i=1}s_i\,{\widetilde\xi}^{\,2}_i\ln \mu^2_i.
\label{PVcoef2}
\eeq
As in \cite{BoxAno}, we find a dependence on the parameters of
Pauli-Villars for the coefficients of the total derivative terms
$\square\Phi^2$ and $\square R$.

One can note that there is an ambiguity in both terms $\cx\Phi^2$
and $\cx R$, related to the choice of $\xi_i$ in the scalar
regulators. The point is that we can choose
$\widetilde{\xi}_i=\mu^2_i$ like it is done in the relation
(\ref{sol3}) or alternatively, take $\widetilde{\xi}_i\equiv 0$,
that also solves the system of equations  \eqref{PVcond1},
\eqref{PVcond2} and  \eqref{PVcond3}.  In the latter case, we recover
the standard value for those anomaly terms which are common for all
regularizations that do not explicitly break conformal invariance.

The remaining question is whether we can remove all the
divergences (\ref{Gamma1loop}) in the general theory by using
the Pauli-Villars regulators. This cannot be done with the
regulators of only scalar type, because those do not
cancel the $X_c$ term that does not appear in the formula
(\ref{Gsc}). Thus, we need to add the regulators of
another type, producing the kinetic scalar counterterm.

Thus, let us introduce spinor regulators with indefinite Grassmann
parity, as suggested in \cite{BoxAno}. The operator which we put
in the bilinear form is
\beq
{\mathcal O}_{j} \,=\,
\hat{H}_j\hat{H}_j^*,
\qquad
%% red
%%\textcolor{red}{
\hat{H}_{j} = \ga^\mu\na_\mu - i\ph_j ,
\qquad
\hat{H}_j^{*}= \ga^\nu\na_\nu  + i\ph_j,
\label{H fermion}
\eeq
where $\ph_j = \tilde{m}_j + h\Phi$, also $\tilde{m}_j = \nu_jM$ is
the mass of the regulator and $\Phi$ is the background scalar (with
generalizations to a multiscalar case described in Sec.~\ref{sec2}).
It is easy to get
\beq
{\mathcal O}_{j}
\,=\, \Box  + i\ga^\al\ph_{j,\al}
- \frac{1}{4}\,R
+ \ph_j^2\,.
%% +\Big(\chi_i -\frac{1}{4}\Big)\hat{1}R,
\label{op two i}
\eeq

The analog of (\ref{Gsc}) in the described fermionic case has the
form \cite{bexi,BoxAno}
\beq
&&
{\bar \Ga}^{(1)}_{j}
\,=\,
\frac{1}{(4\pi)^2}\,\int d^4x \,\sqrt{g}\,
\Big\{ - \tilde{m}_j^4\Big(\frac{1}{\ep}+\frac32\Big)
+ \Big( \frac{1}{6}\,R - 2h^2\Phi^2\Big)\tilde{m}_j^2
\Big(\frac{1}{\ep}+1\Big)
\nonumber
\\
&&
\qquad
\quad
+\,\,
\frac14\,C_{\mu\nu\al\be}
\Big[\frac{1}{10\,\ep}+k^f_W(a)\Big] C^{\mu\nu\al\be}
+ \frac12h^2\,\big(\na_\al\Phi \big)
\Big(\frac{2}{\ep} + 4A\Big)\,\big(\na^\al\Phi\big)
\nonumber
\\
&&
\qquad
\quad
+\,\,
\frac12\,h^2\Phi^2
\Big(\frac{1}{3\ep}+\frac29+\frac{8A}{3a^2}\Big)R
- \frac12\,h^4\Phi^2 \Big(\frac{2}{\ep}+4A\Big)
\Phi^2
\,+\, \frac12\,R\,k^f_R(a)\,R
\Big\}\,.
\label{Gaf}
\eeq
The finite nonlocal form factors $k^f_W$ and $k^f_R$ can be found
in the second reference in \cite{apco} and we used notations
\beq
A\,=\,1-\frac{1}{a}
%% red
%%\textcolor{red}{
\ln\Big(\frac{1+a/2}{1-a/2}\Big)\,,
\qquad
a^2 = \frac{4\Box}{\Box - 4m^2}.
\label{Aa}
\eeq
The cancelation of the vacuum $N_{1/2}$-dependent part of
(\ref{Gamma1loop}) can be done using regulators (\ref{op two i})
and the action
\beq
&&
{\bar\Ga}^{(1)}_{reg,f}
\,=\,
\sum^{N_f}_{j=0}t_j{\bar\Ga}^{(1)}_{j}(\tilde{m}_j),
\label{EA_reg-fer}
\eeq
where $j=0$ correspond to each of the physical fermions. This
means, we shall need $N_{1/2}$ copies of each of these
regulators, exactly as we need $N_0$ copies in the scalar case.
The conditions for the coefficients $t_j$ and $\nu_j$ are completely
similar to \eqref{PVcond1}, \eqref{PVcond2}, and \eqref{PVcond3}
if we assume $N_f=5$ and
replace $s_i \to t_j$ and  $\mu_i \to \nu_j$, this time with
$\widetilde{\xi}_j \equiv 0$. Correspondingly, the solution is
given by (\ref{sol1}) and  (\ref{sol2}) with the same replacements.

At this point we can make two relevant observations. The first is
that the cancelation of the $\Phi$-dependent terms in
(\ref{Gamma1loop})
can be provided by the tuning of the artificial couplings $\ka$ and
$h$ in the regulator actions (\ref{EA_reg}) and  (\ref{EA_reg-fer}).
This can be provided for any scalar contents and, correspondingly,
for any $\be_\la$ and $\ga_\Phi$, even regardless of the possible
gauge fixing ambiguity in the last quantity. Thus, we do not need
to worry about the dependence on the scalar $\Phi$ in the rest of
our consideration.

The second point is as follows. Looking at the expressions
%% (\ref{Gamma1loop}),  (\ref{Gaf}),
(\ref{w}), (\ref{b}), and
(\ref{c}) it is clear that the cancelation of divergences for a
general set $N_{0,\,1/2,\,1}$ cannot be done only with scalar
and fermion regulators. Thus, we need to introduce vector
regulators to take care about $N_1$-dependent part. This
regulator do not need to depend on $\Phi$ owing to the
first observation. Here we meet an aparent problem because
the regulator should be massive and the massless limit in the
gravitational form factors are known to manifest discontinuity
\cite{BuGui}. However,  the solution of this problem is known
for a long time \cite{Slavnov-PV,AsoFalc}. The regulator fields
are unphysical and do not need to satisfy the same conditions of
consistency as physical fields, such as correspondence between
spin and statistics, or the absence of the unphysical modes.
Thus, we can define the regulators for the vector part in the
form
\beq
&&
{\bar\Ga}^{(1)}_{reg,v}
\,=\,
\sum^{N_v}_{k=0}r_k {\bar\Ga}^{(1)}_{k}(\check{m}_k),
\label{EA_reg-vec}
\eeq
where
\beq
{\bar\Ga}^{(1)}_{k}(\check{m}_k)
\,=\,\frac{i}{2} \Tr \Ln \big( \de^\al_{\,\be}\cx - R^\al_{\,\be}
+ \check{m}^2_k \de^\al_{\,\be}\big)
\,-\,i\Tr\Ln \big(\cx + \check{m}^2_k\big)
\label{Gavecmas}
\eeq
and $\check{m}^2_k = \rho_k M$ is the mass of the regulator.
It is important that in (\ref{Gavecmas}) we have double contribution
of the scalar mode, like in the massless Faddeev-Popov ghosts cases.
This is different from the single contribution in case of a massive
vector \cite{bavi85,BuGui}. This feature guarantees the cancelation
of the divergences in the $N_1$ vacuum part of (\ref{Gamma1loop})
if we chose the conditions similar to \eqref{PVcond1}, \eqref{PVcond2},
and \eqref{PVcond3} for the coefficients $r_k$ and $\rho_k$, take
$N_v=5$ and replace $s_i \to r_k$ and  $\mu_i \to \rho_k$, of
course with 
%% red
%%\textcolor{red}{
$\widetilde{\xi}_k \equiv 0$, as in the fermion case.
The solution is (\ref{sol1}) and  (\ref{sol2}) with the appropriate
replacements. The explicit form of (\ref{Gavecmas}), with the full
set of nonlocal form factors, can be found in the second reference
in \cite{apco} or easily extracted from \cite{BuGui}.

Thus, we have proven that the cancelation of divergences is
possible for any particle contents and beta functions of the
underlying model. The ambiguity is present and is given by
(\ref{anomaly_amb}).
%%  red
%%\textcolor{red}{
Let us stress that, despite the local terms satisfy
the power counting in the theory, the ambiguity described above
cannot be fixed by the change of renormalization parameter $\mu$
because the corresponding terms are not present in the initial
classical action and are not subject of the UV renormalization.
This aspect of the theory was previously discussed in the
classical paper \cite{Brown-Collins_1980} and more recently
in \cite{BoxAno}.
The ambiguity in the local terms can be fixed only by introducing
these nonconformal terms from the very beginning with arbitrary
coefficients, that can be fixed by experiment and not by a
particular regularization scheme.

%%%%%%%%%%%%%%%%%%%%%%%%%%%%%%%%%%
%%%%%%%%%%%%%%%%%%%%%%%%%%%%%%%%%%
%%%%%%%%%%%%%%%%%%%%%%%%%%%%%%%%%%
\section{Anomaly-induced effective action in the IR}
\label{sec5}

In this section, we explore the anomaly-induced action at low
energies, i.e. in the IR. The scheme which we shall use is partially
similar to the one presented in \cite{MotVaul,GianMott09} for
the electromagnetic and metric background. In what follows, we
shall see that the scalar case provides interesting novelties and
unexpected connection to other approaches.

For our purposes, the most useful version of the effective action
is not (\ref{aux_fields}) but the nonlocal form (\ref{nonlocal}).
Then, our approximations are as follows.

$i)$ All matter fields are (at least approximately) massless and all
$\xi_i\approx \frac16$, such that the conformal symmetry holds at
the classical level. In this case, the anomaly-induced action is a
good-quality approximation.

$ii)$ Scalar terms $ \Phi^4$ and $X_c$ %% (\na\Phi)^2+\xi R\Phi^2$
are dominating over the curvature terms, i.e.,
\beq
\big| \Phi^2\big|  \gg \big|R_{\ldotp\ldotp\ldotp\ldotp}\big|
\qquad
\mbox{and}
\qquad
\big|(\nabla\Phi)^2\big| \gg \big|R^2_{\ldotp\ldotp\ldotp\ldotp}\big|
\label{IR-1}
\eeq
for all the components of the curvature tensor
$\,R_{\ldotp\ldotp\ldotp\ldotp}$.

$iii)$ As always in general relativity, the IR limit means that the
gravitational field is weak.\footnote{
%%   red
%%\textcolor{red}{
This means that in GR the
low-energy approximation is actually a weak curvature limit.
%%%%%   end of red
}
According to this assumption, the
dominating metric-dependent quantities are those which do not
vanish in the linear order in the metric perturbations $h_{\mu\nu}$
over the flat background. In particular, this implies
$|\cx R| \gg |R^2_{\ldotp\ldotp\ldotp\ldotp}|$ for all curvature
contractions.
Thus, the scalar-dependent terms and $\cx R$-terms represent the
most relevant part of the anomaly-induced effective action.
In general, the anomaly in the vacuum sector is a sub-leading part
in the given approximation and has to be treated as auxiliary element
to arrive at the scalar-dependent contributions.

The non-local structures in the induced action reduce to a more
simple form because
\beq
G=\Delta^{-1}_{4}=\Big(\square^2
+ 2R^{\mu\nu}\nabla_{\mu}\nabla_{\nu}
- \dfrac{2}{3}R\square
+ \dfrac{1}{3}R^{;\mu} \nabla_{\mu}\Big)^{-1}
\,\approx \,\frac{1}{\square^2}.
\label{apprG}
\eeq
The leading terms in the expression (\ref{Y}) are those
with $\Phi$ and $\cx R$, hence
\beq
&&
E_4 - \frac{2}{3} \cx R + \frac{1}{b}Y
\,\,\approx \,\,
-\,\frac23\,\cx R
\,-\,
\frac{1}{b}\Big(\ga_\Phi X_c- \frac{1}{4!}\,\widetilde{\be}_\la \Phi^4 \Big),
\label{appr-int}
\eeq
where we %% denoted
%% $X_c \equiv \frac{1}{2}\Big[(\na \Phi)^2 + \frac{1}{6}R\Phi^2\Big]$,
splitted $Y$ into two parts according to (\ref{Y}).

After a small algebra, the nonlocal part of expression
\eqref{nonlocal} boils down\footnote{On top of this, there is a
usual local part, i.e., \  $\displaystyle\,-
\int_x\,\Big(\frac{2 b+3c}{36}R^2
+ \frac{\be_\tau}{6} R\Phi^2\Big).$}  to become
\beq
&&
\Gamma_{ind,nonloc} \,\,\, \approx \,\,\, \frac{b}{18}\int_xR^2
\,\,-\,\,
\frac{1}{6}\int_x\int_y\Big(
\frac{1}{4!}\,\widetilde{\be}_\la \Phi^4
- \gamma_\Phi X_c\Big)_x\,\,
\Big(\frac{1}{\cx^2}\Big)_{x,y}\,\, \big(\cx R \big)_y
\nn
\\
&&
\qquad
\qquad
\quad
\,\,=\, \frac{b}{18}\int_xR^2
\,\,-\,\,
\frac{1}{6}\int_x\int_y\Big(
\frac{1}{4!}\,\widetilde{\be}_\la \Phi^4
- \gamma_\Phi X_c\Big)_x\,\,
\Big(\frac{1}{\cx}\Big)_{x,y}\,\, \big(R \big)_y.
\label{appr}
\eeq
The last formula is interesting and deserves a few observations.

$ i)$ \ The first integral in this expression shows that there is a
%% red
%%\textcolor{red}{
modified coefficient of the local $R^2$-term in the IR.
Let us note that this modification is a direct consequence of that
the $R^2$-term was hidden the first line of Eq.~(\ref{nonlocal})
under the flat space limit. In this way we just recover the $b$
term in the induced anomaly action in that limit.
According to (\ref{b}), the addition to the coefficient of
$R^2$ has the value $b/18$, hence it is negative and its
magnitude is about $0.01-1$. The effect is insufficient to explain
the huge coefficient (about $5\times 10^8 $) of the $R^2$-term in
the Starobinsky model \cite{star,star83} and we still have the
challenging problem of deriving this coefficient, as discussed
recently in \cite{StabInstab}.
%%%  end of red

%% red
%%\textcolor{red}{
It is worth noting that the classical $R^2$ term required to explain
the observations in the inflationary model \cite{star,star83} leads
to the propagating scalar mode of the metric in the flat-space limit,
exactly as it is the case for Einstein's gravity corrected by anomaly
\cite{Mottola-2017}. Anyway, it is quite interesting, from the
theoretical side, that the quantum effects may enlarge the spectrum
of the gravitational waves in the initially conformal theory, where
the scalar mode is absent.
%%%  end of red

$ii)$ \ In the symmetry-breaking regime, we have
$\Phi \approx v$ i.e., the scalar field is approximately a constant.
Then the second integral in (\ref{appr}) takes the form
\beq
&&
-\frac{1}{6}\int_x\int_y
\Big(\frac{1}{4!}\, \widetilde{\be}_{\la}v^4
\,-\, \frac16\,\gamma_\Phi Rv^2\Big)_x \,
\Big(\frac{1}{\cx}\Big)_{x,y} R(y).
\label{SSB-term1}
\eeq
In the last summand we meet the first order term $R\,\square^{-1}R$
of the known nonlocal expression \cite{DW-07}. Physically, this
may be the case for the Standard Model Higgs scalar, at the energies
greater than the typical value $v=246$ GeV. At this scale, the anomaly
is an appropriate approximation and, compared to the Planck energy
or the GUT scale, this may be an IR region.

Let us note that the terms $R\,\square^{-1}R$, \
$R\,(\square^{-1}R)^2$, etc,  have the same
global scaling as the usual $R$-term. For this reason, including
these terms  with the properly chosen coefficients may produce
a small deviation from GR in the cosmological setting. In particular,
these terms may serve as a replacement of a small cosmological
constant in the phenomenological description of the acceleration
of the Universe. Remarkably, we can arrive at these terms starting
from the conformal anomaly and taking the low-energy
approximation as described above.

It is remarkable that even in the next orders of the IR
approximation, we do not meet the terms ${\mathcal O}(Y^2)$
because these structures don't appear in the induced action
(\ref{nonlocal}). As a
consequence, the anomaly-induced action does not generate
$R \cx^{-2}R$-type terms. It is known that these terms cannot
be obtained, also, from the spontaneous symmetry breaking by the
curvature expansion \cite{Sponta}. The interest to derive such a
structure
is because it is used for the phenomenological description of the
accelerated expansion of the Universe \cite{Magg2} (see also
references therein), replacing the cosmological term.  However,
it is not easy to obtain it from the quantum considerations, at
least at the one-loop order.

$iii)$ \
Let us come back to the formula \eqref{appr} and
consider $g_{\mu\nu}=\bar{g}_{\mu\nu}e^{2\sigma}$ and
$\Phi=\bar{\Phi}e^{-\sigma}$. Assuming a weak gravitational
field, in the linear in $\sigma$ approximation we get
\beq
&&
\frac{1}{\cx}\,=\,e^{2\sigma}\frac{1}{\bar{\cx}},
\qquad
R\,=\,e^{-2\si}\big[\bar{R}-6\bar{\cx}\sigma+O(\sigma^2)\big],
\label{R-exp}
\eeq
where
$\bar{\cx}=\bar{g}^{\mu\nu}\partial_\mu\partial_\nu$.
Then, after integration with the delta function, we arrive at
\beq
&&
\Ga_{2}
\,\,=\,\,
\int_x %% \int_y
\Big(\frac{1}{4!}\,\widetilde{\be}_{\la} \bar{\Phi}^4
\,-\, \gamma_\Phi \bar{X}_c
\Big)\sigma,
\label{appr1}
\eeq
with a local expression in the integrand. This output is rather
natural, but still amusing. It confirms explicitly that the
anomaly-induced action is, in fact, a local version of the
renormalization group corrected classical action. The usual
constant scaling parameter in curved space \cite{tmf84,book}
is replaced by the local function $\si$, making curved
spacetime renormalization group closer to applications
\cite{PoImpo}.

Regarding (\ref{appr1}) as a loop correction to the classical
action of scalar field, the effective action of scalar field can
be obtained by the changes
\beq
X_c \longrightarrow \bar{X}_c(1+\gamma_\Phi \sigma),
\qquad
\lambda\Phi^4
\longrightarrow
\bar{\Phi}^{4}(\lambda+\widetilde{\beta}_\lambda \sigma)
\label{RGimprove}
\eeq
as it should be under the renormalization group - based improvement.
In particular, one can use this approach to recover the one-loop
effective potential. For this, one has to follow the procedure:
\beq
&&
\bar{\Phi} \,\longrightarrow\, \Phi,
\qquad
\sigma \,\longrightarrow \,\ln \frac{\Phi}{\,\,\Phi_0},
\label{param}
\eeq
that is coherent with the scalar part of (\ref{conf-4}). At the end, one
has to replace $\bar{\Phi}$ by $\Phi$ and $\bar{g}_{\mu\nu}$ by
$g_{\mu\nu}$, as we did in the previous sections.

For the one-loop effective potential, in this way we arrive at
\beq
V_{eff}^{(1)}
\,\,=\,\,
%% \textstyle
\frac{1}{4!}\,\Big(
\lambda\,+\,\frac12\,\widetilde{\beta}_\la
\ln  \frac{\Phi^2}{\mu^2} \Big) \Phi^4
\,\,-\,\,
%% \displaystyle
\frac{1}{12}
\Big( 1\,+\,  %% \frac12\,
\ga_\Phi \ln %% \textstyle
\frac{\Phi^2}{\mu^2}\Big)R\Phi^2,
\label{EfPo}
\eeq
where $\widetilde{\be}_{\la} =\be_\la + 4\la\gamma_\Phi$.
Formula (\ref{EfPo}) requires only the renormalization conditions
to become the standard expression for the massless conformal
theory in curved spacetime \cite{book}, that is a direct extension
of the Coleman and Weinberg potential \cite{ColeWein}. One can
include $X_c$ term together with $R\Phi^2$ and get the first term
in the derivative expansion of effective action. This would be
completely equivalent to the renormalization-group based method
\cite{BuchWolf,book}.
%% red
%%\textcolor{red}{
This equivalence may be observed in the
scattering of gravitational perturbations on a flat background,
as it was recently considered in \cite{Mottola-2017}. The
relation between the anomaly-induced action and the
effective potential (\ref{EfPo}) shows that this equivalence
can be extended to other applications.
%%%  end of red

The procedure of
deriving the effective potential and other terms in the effective
action from anomaly is expected to work only for those terms which
have UV logarithmic divergences and are not related to the masses
of quantum fields. The anomaly picks up the UV sector of quantum
corrections and it is not a surprise that, with certain skills, one can
recover the corresponding terms, e.g. in the potential.

%%%%%%%%%%%%%%%%%%%%%%%%%%%%%%%%%%
%%%%%%%%%%%%%%%%%%%%%%%%%%%%%%%%%%
%%%%%%%%%%%%%%%%%%%%%%%%%%%%%%%%%%
\section{Conclusions and discussions}
\label{sec6}

We have formulated the anomaly-induced effective action in curved
spacetime with additional background scalars. The result is given in
the covariant nonlocal form (\ref{quantum1}), it is also presented as
local action (\ref{aux_fields}) using two auxiliary scalars. The
prescription for deriving the induced effective actions is pretty
well-known, including a very general form \cite{6d} and enables one
to add extra fields, as we did with scalars. Let us note that the
non-covariant form of induced metric-scalar action has been
previously obtained in \cite{shocom,asta}, including for light
massive quantum fields.  We hope that the covariant forms obtained
above can be also extended to the light massive fields case, but at
the moment this remains an unsolved problem.

The most complicated new results of the present work are the
extension of the Pauli-Villars regularization ambiguities for the
local terms in the effective action to an arbitrary quantum theory
including scalar fields and, especially, the consideration of the
low-energy limit in the induced effective action.

The problem of ambiguities has been previously addressed in
\cite{anomaly-2004,BoxAno} for pure gravity and for the case
of a single real scalar field. The extension of  this analysis to an
arbitrary gauge model required a new type of fine tuning for the
scalar and spinor regulators, in the part of their coupling to
background scalars, and also introducing massive vector field
regulators in such a way that the massless limit does not show
a usual discontinuity. We have described how this can be done in
Sec.~\ref{sec3}. The main output of this part of our work is that
there is always an ambiguity in the one-loop finite quantum
correction of the $R\Phi^2$-type.

The main question related to the conformal symmetry and anomaly
is whether the general structure of anomaly or, equivalently, the
form of induced action (\ref{quantum1}) can be preserved beyond
the one-loop level. Our results show that, starting from the two-loop
approximation, preserving the structure of anomaly requires a
fine tuning of the Pauli-Villars regulators, such that the combination
$\be_\tau + 3\rho$ i.e., the coefficient of $\cx \Phi^2$ term in the
anomaly (\ref{anomaly_amb}), vanish. Since the only instrument
to provide this cancelation is the choice of $\widetilde{\xi}_i = 0$ or
$\widetilde{\xi}_i = \mu_i^2$ in (\ref{PVcoef1}), this cancelation
looks impossible.

%% \textcolor{red}{\LARGE \textbullet}
%%  red
What happens after that can be described as follows. The non-local
anomalous terms do not produce UV divergences. In contrast, the
local $R\Phi^2$ term does. The one-loop $R\Phi^2$ looks exactly
as the classical non-conformal term with an extra $\hbar$. Then, for
the second (superficial) integration in the two-loop diagrams this
term produces the ${\mathcal O}(\hbar^2)$ term with the $1/\ep$
divergence. Consequently, the two-loop beta-function for $\xi$ has
no fixed point at $\xi=\frac16$ and the whole conformal framework
breaks down.
Taking the described situation into account, at the two-loop order
of the theory with scalars we have to deal with $\xi \neq \frac16$
and the theory becomes essentially non-conformal. Thus, the general
expectation is that beyond the first loop, the structure of anomaly
and the anomaly-induced action becomes qualitatively different and
the conformal symmetry can be regarded only as an approximation.
It is worth mentioning that this scenario is known in
the literature both from the direct higher-loop calculations
\cite{Hathrell} in the $\phi^4$ theory and, most important, from
the general analysis of the conformal Ward identities \cite{tmf84}.
In this work, we clarified the role of regularization in this scenario.
The last observation in this part is that, qualitatively the same
situation takes place with the term $\cx R$ in all possible versions
of conformal quantum gravity theory.

%% {\LARGE \textbullet}

Our analysis of the low-energy limit in the anomaly-induced action
has been performed in the way regarded standard in general relativity.
This approach led us to the reduced form (\ref{appr}) of the action.
This expression manifested a few qualitatively new properties. First
of all, we have found a new contribution to the local $R^2$ term
in the IR. Second, in the IR we can reproduce, starting from the
induced action, the non-local structure $R\cx^{-1}R$ introduced
earlier in the literature \cite{apco}, especially used in cosmology
on the purely phenomenological \textit{ad hoc} basis  \cite{DW-07}.
On the other hand, it is not possible to generate $R\cx^{-2}R$ term,
also existing in the literature on modified gravity \cite{Magg2}. Third,
we confirmed that the anomaly-induced
action represents a useful local version of the renormalization group
in curved spacetime. In particular, it turns out possible to recover
the effective potential of the conformal scalar field starting from
the induced action.

%%%%%%%%%%%%%%%%%%%%%%%%%%%%%%%%
%%%%%%%%%%%%%%%%%%%%%%%%%%%%%%%%
\section*{Acknowledgements}
%% \label{secAck}

W.C.S. and P.R.B.V. are grateful to CAPES for supporting their Ph.D.
projects.
The work of M.A. is partially supported by Spanish MINECO/FEDER
grant PGC2018-095328-B-I00 and DGA-FSE grant 2020-E21-17R.
I.Sh. is partially supported by Conselho Nacional de Desenvolvimento
Cient\'{i}fico e Tecnol\'{o}gico - CNPq (Brazil) under the grant
303635/2018-5 and by Funda\c{c}\~{a}o de Amparo \`a Pesquisa de
Minas Gerais - FAPEMIG, under the project PPM-00604-18.

%%%%%%%%%%%%%%%%%%%%%%%%%%%%%%%%%%
%%%%%%%%%%%%%%%%%%%%%%%%%%%%%%%%%%
%%%%%%%%%%%%%%%%%%%%%%%%%%%%%%%%%%

%%%%%%%%%%%%%%%%%%%%%%%%%%%%%%
%% \bibliographystyle{unsrturl}
%\bibliography{References}

\begin{thebibliography}{99}
		
%\bibitem{DeWitt65} B.S. DeWitt,
%\textit{ Dynamical Theory of Groups and Fields} (Gordon and Breach, New York U.S.A, 1965).
\small{

\bibitem{CapDuf-74} D. M. Capper, M. J. Duff and L. Halpern,
\textit{ Photon corrections to the graviton propagator,}
Phys. Rev. {\bf D10} (1974) 461;
\
D. M. Capper and M. J. Duff,
\textit{ Neutrino corrections to the graviton propagator,}
Nucl. Phys. {\bf B82} (1974) 147.

\bibitem{duff77} M.J. Duff,
\textit{ Observations On Conformal Anomalies,}
Nucl.Phys. {\bf B125} (1977) 334.
%% DOI: 10.1016/0550-3213(77)90410-2
		
\bibitem{ddi} S. Deser, M.J. Duff and C. Isham,
\textit{ Nonlocal conformal anomalies},
Nucl. Phys. {\bf B111} (1976) 45.
%% DOI: 10.1016/0550-3213(76)90480-6
		
\bibitem{duff94} M.J. Duff,
\textit{ Twenty years of the Weyl anomaly,}
Class. Quant. Grav. {\bf 11} (1994) 1387,
%% DOI: 10.1088/0264-9381/11/6/004
hep-th/9308075.
		
\bibitem{rie} R.J. Riegert,
\textit{ A non-local action for the trace anomaly},
Phys. Lett. {\bf B134} (1984) 56. %% -60
%% DOI: 10.1016/0370-2693(84)90983-3
		
\bibitem{frts84} E.S. Fradkin and A.A. Tseytlin,
\textit{ Conformal anomaly in Weyl theory and anomaly free
superconformal theories,}
Phys. Lett. {\bf B134} (1984) 187.
%% DOI: 10.1016/0370-2693(84)90668-3

\bibitem{AntMot92} I.~Antoniadis and E.~Mottola,
\textit{$4-D$ quantum gravity in the conformal sector,}
Phys. Rev. \textbf{D45} (1992) 2013. %% -2025
%% doi:10.1103/PhysRevD.45.2013
		
\bibitem{PoImpo} I.L.~Shapiro,
\textit{ Effective action of vacuum: semiclassical approach},
Class. Quant. Grav. {\bf 25} (2008) 103001,
%% DOI: 10.1088/0264-9381/25/10/103001
arXiv:0801.0216.
%%  \bibitem{conf06} I.L. Shapiro,
%%  \textit{ Local conformal symmetry and its fate at quantum level,} arXiv:hep-th/0610168;
%%  Talk presented at the Fifth International Conference on Mathematical Methods in Physics, Rio de %%   Janeiro, Brazil. PoS \textbf{IC2006} (2006) 030.
\,
%% red
%%\textcolor{red}{
\bibitem{Bardeen-1995} W.A.~Bardeen,
\textit{On naturalness in the standard model,}
FERMILAB-CONF-95-391-T.
\,
\bibitem{Meissner-2007}
K.A.~Meissner and H.~Nicolai,
\textit{Conformal symmetry and the Standard Model,}
Phys. Lett. \textbf{B648} (2007)  312,   %%   -317
%%  doi:10.1016/j.physletb.2007.03.023
hep-th/0612165.
%%%  end of red

\bibitem{Higgsmass12} J.~Elias-Miro, J.R.~Espinosa, G.F.~Giudice,
G. Isidori, A. Riotto and A.~Strumia,
\textit{Higgs mass implications on the stability of the electroweak vacuum,}
Phys. Lett. {\bf B709} (2012) 222,
%% DOI: 10.1016/j.physletb.2012.02.013
arXiv:1112.3022.
		
\bibitem{PlanckH12} M. Holthausen, K.S. Lim and M. Lindner,
\textit{Planck scale boundary conditions and the Higgs mass,}
JHEP {\bf 02} (2012) 037,
%% DOI: 10.1007/JHEP02(2012)037
arXiv:1112.2415.
		
\bibitem{stabV12}
J.~Elias-Miro, J.R.~Espinosa, G.F.~Giudice, H.M.~Lee and A.~Strumia,
\textit{Stabilization of the Electroweak Vacuum by a Scalar Threshold Effect,}
JHEP {\bf 06} (2012) 031,
%% DOI: 10.1007/JHEP06(2012)031
arXiv:1203.0237.

\bibitem{DrumSho79} I.T. Drummond and G.M. Shore,
\textit{Conformal anomalies for interacting scalar fields in curved spacetime,}
Phys. Rev. {\bf D19} (1979) 1134.
%% DOI: 10.1103/PhysRevD.19.1134
		
\bibitem{Hathrell} S.J. Hathrell,
\textit{ Trace anomalies and $ \la\ph^{4} $ theory in curved space,}
Annals of Phys. {\bf 139} (1982) 136.
%% DOI: 10.1016/0003-4916(82)90008-2
		
\bibitem{ValGoni87} M.A. Valle and M.A. Goni,
\textit{ Conformal transformation of the one-loop effective action for $ \la\ph^{4} $ theory,}
Phys. Rev. {\bf D36} (1987) 615.
%% DOI: 10.1103/PhysRevD.36.615
		
\bibitem{AlvBarc} M.S. Alves and J. Barcelos-Neto,
\textit{Path Integrals and the Trace Anomaly in a Self-Interacting Scalar Theory,}
Mod. Phys. Lett. {\bf A4} (1989) 115.
%% DOI: 10.1142/S0217732389000216

\bibitem{anomaly-2004} M. Asorey, E.V. Gorbar and I.L. Shapiro,
\textit{Universality and ambiguities of the conformal anomaly,}
Class. Quant. Grav. {\bf 21} (2004) 163,
%% DOI: 10.1088/0264-9381/21/1/011
hep-th/0307187.

\bibitem{BoxAno}
M. Asorey, G. de Berredo-Peixoto and I.L. Shapiro,
\textit{Renormalization ambiguities and conformal anomaly
in metric-scalar backgrounds,}
Phys. Rev. {\bf D74} (2006) 124011,
%% DOI: 10.1103/PhysRevD.74.124011
arXiv:hep-th/0609138.

\bibitem{book} I.L. Buchbinder, S.D. Odintsov and I.L. Shapiro,
\textit{Effective Action in Quantum Gravity}
(IOP Publishing, Bristol, 1992).

\bibitem{OUP} I.L. Buchbinder and I.L. Shapiro,
%%  \bibitem{QFT-OUP} I.L. Buchbinder and I.L. Shapiro,
\textit{Introduction to Quantum Field Theory with Applications
to Quantum Gravity} (Oxford University Press, 2021).

\bibitem{tmf84} I.L. Buchbinder,
\textit{ On Renormalization group equations in curved space-time,}
Theor. Math. Phys. {\bf 61} (1984) 393.
%% \bibitem{Buch-TMF/84} I.L. Buchbinder,
%% DOI: 10.1007/BF01035006

\bibitem{birdav} N.D. Birrell and P.C.W. Davies,
\textit{Quantum fields in curved space,}
(Cambridge University Press, Cambridge, 1982).

\bibitem{VoTyu76} B.L.~Voronov, I.V.~Tyutin,
\textit{Models of asymptotically free massive fields,}
Yad.Fiz. \textbf{23} (1976) 664. %% -675

\bibitem{GUT-90} I.L. Buchbinder, I.L. Shapiro and E.G. Yagunov,
\textit{ The asymptotically free and asymptotically conformal
invariant Grand Unification theories in curved space-time,}
Mod. Phys. Lett. {\bf A5} (1990) 1599,
%% DOI: 10.1142/S0217732390001827

\bibitem{2Yoon} Y. Yoon and Y. Yoon,
\textit{ Asymptotic conformal invariance of SU(2) and standard
models in curved space-time,}
Int. J. Mod. Phys. {\bf A12} (1997) 2903,
%% DOI: 10.1142/S0217751X97001602
hep-th/9612001.
		
\bibitem{DeserSchwimmer}
S.~Deser and A.~Schwimmer,
{\it Geometric classification of conformal anomalies in
arbitrary dimensions,}
Phys. Lett.  \textbf{B309} (1993) 279, %% -284
%% doi:10.1016/0370-2693(93)90934-A
hep-th/9302047.

\bibitem{6d} F.M. Ferreira and I.L. Shapiro,
{\it Integration of trace anomaly in $6D$,}
Phys. Lett. {\bf B772} (2017) 174, % -178.
%% DOI: 10.1016/j.physletb.2017.06.014
arXiv:1702.06892.

\bibitem{polyakov81} A.M. Polyakov,
{\it Quantum geometry of bosonic strings,}
%% DOI: 10.1016/0370-2693(81)90743-7
Phys. Lett. {\bf B207} (1981) 211.

\bibitem{BuchOdShp85} I.L. Buchbinder, S.D. Odintsov
and I.L. Shapiro,
\textit{ Nonsingular cosmological model with torsion induced by vacuum
quantum effects,}
Phys. Lett. {\bf B162} (1985) 92.
%% DOI: 10.1016/0370-2693(85)91067-6
		
\bibitem{anhesh} J.A. Helayel-Neto, A. Penna-Firme and I.L. Shapiro,
\textit{ Conformal symmetry, anomaly and effective action for
metric-scalar gravity with torsion,}
Phys. Lett. {\bf B479} (2000) 411,
%% DOI: 10.1016/S0370-2693(00)00342-7
gr-qc/9907081.
		
\bibitem{MaurShp15} S. Mauro and I.L. Shapiro,
\textit{ Anomaly-induced effective action and Chern-Simons
modification of general relativity,}
Phys. Lett. {\bf B746} (2015) 372,
%% DOI: 10.1016/j.physletb.2015.05.045
arXiv:1412.5002.

\bibitem{FrTs-superconf} E.S. Fradkin and A.A. Tseytlin,
\textit{ Asymptotic freedom on extended conformal supergravities,}
Phys. Lett. {\bf B110} (1982) 117; %% -122
%% DOI: 10.1016/0370-2693(82)91018-8
\textit{ One-loop beta function in conformal supergravities,}
Nucl. Phys. {\bf B203} (1982) 157. %% -178
%% DOI: 10.1016/0550-3213(82)90481-3
		
\bibitem{Paneitz} S. Paneitz,
\textit{ A quartic conformally covariant differential operator for
arbitrary pseudo Riemannian manifolds,}
MIT preprint - 1983; SIGMA {\bf 4} (2008) 036,
%% DOI: 10.3842/SIGMA.2008.036
arXiv:0803.4331.


\bibitem{a} I.L. Shapiro and A.G. Jacksenaev,
\textit{ Gauge dependence in higher derivative quantum gravity and
the conformal anomaly problem,}
Phys. Lett. {\bf B324} (1994) 286.
%% DOI: 10.1016/0370-2693(94)90195-3
		
\bibitem{MazMott01} P.O. Mazur and E. Mottola,
\textit{ Weyl cohomology and the effective action for conformal
anomalies,}
Phys. Rev. {\bf D64} (2001) 104022,
%% DOI: 10.1103/PhysRevD.64.104022
hep-th/0106151.
		
\,
%% red
%%\textcolor{red}{
\bibitem{Mottola-2017} E.~Mottola,
\textit{Scalar gravitational waves in the effective theory
of gravity,}
JHEP \textbf{07} (2017) 043;
Erratum: JHEP \textbf{09} (2017) 107,
%%   doi:10.1007/JHEP07(2017)043
arXiv:1606.09220. %%  [gr-qc]].
%%%  end of red

\bibitem{GianMott09} M. Giannotti and E. Mottola,
\textit{ Trace anomaly and massless scalar degrees of freedom in gravity,}
Phys. Rev. {\bf D79} (2009) 045014,
%% DOI: 10.1103/PhysRevD.79.045014
arXiv:0812.0351.

\bibitem{shocom} I.L. Shapiro and Joan Sol\`a,
\textit{ Massive fields temper anomaly-induced inflation:
the clue to graceful exit?,}
Phys. Lett. {\bf B530} (2002) 10,
DOI: 10.1016/S0370-2693(02)01355-2
hep-ph/0104182.
		
\bibitem{asta} A.M. Pelinson, I.L. Shapiro and F.I. Takakura,
\textit{ On the stability of the anomaly-induced inflation,}
Nucl. Phys. {\bf B648} (2003) 417,
%% DOI: 10.1016/S0550-3213(02)00999-9
hep-ph/0208184.

\bibitem{apco} E.V. Gorbar and I.L.~Shapiro,
\textit{ Renormalization group and decoupling in curved space},
JHEP {\bf 02} (2003) 021,
%% DOI: 10.1088/1126-6708/2003/02/021
hep-ph/0210388;
\textit{ Renormalization group and decoupling in curved space,
II. The standard model and beyond},
JHEP {\bf 02} (2003) 021,
%% DOI: 10.1088/1126-6708/2003/06/004
hep-ph/0303124.

\bibitem{bexi} G. de Berredo-Peixoto, E.V. Gorbar and I.L. Shapiro,
\textit{On the renormalization group for the interacting massive
scalar field theory in curved space,}
Class. Quant. Grav. {\bf 21} (2004) 2281,  %%   -2290;
hep-th/0311229.

\bibitem{Slavnov-PV} A.A. Slavnov,
\textit{The Pauli-Villars regularization for nonabelian gauge theories,}
Theor. Math. Phys. {\bf 33} (1977) 210;
\
T.D. Bakeyev and A.A. Slavnov,
\textit{Higher covariant derivative regularization revisited,}
Mod. Phys. Lett. {\bf A11} (1996) 1539, hep-th/9601092.

\bibitem{AsoFalc} M. Asorey and F. Falceto,
\textit{Geometric regularization of gauge theories,}
Nucl.Phys. {\bf B327}(1989)427.
%% ; Phys. Rev. {\bf D54} (1996) 5290.
	
\bibitem{bavi90} A.O. Barvinsky and G.A. Vilkovisky,
\textit{ Covariant perturbation theory (II). Second order in the
curvature. General algorithms,}
Nucl. Phys. {\bf B333} (1990) 471.
%% DOI: 10.1016/0550-3213(90)90047-H

\bibitem{Avra89} I.G. Avramidi,
\textit{ Covariant methods of study on the nonlocal structure of effective action,}
Yad. Fiz. (Sov. Journ. Nucl. Phys.) {\bf 49} (1989) 1185.

\bibitem{Omar4D} S.A. Franchino-Vi\~nas, T. de Paula Netto,
I.L. Shapiro and O. Zanusso,
\textit{ Form factors and decoupling of matter fields in four-dimensional
gravity,}
Phys. Lett. {\bf B790} (2019) 229, %% -236
%%%%    DOI: 10.1016/j.physletb.2019.01.021
arXiv:1812.00460.

%% red
%%\textcolor{red}{
\bibitem{Meissner-2008} K.A.~Meissner and H.~Nicolai,
\textit{Effective action, conformal anomaly and the
issue of quadratic divergences,}
Phys. Lett. \textbf{B660} (2008)  260,   %%   -266
%%   doi:10.1016/j.physletb.2007.12.035
hep-th/0710.2840.
%%%  end of red

\bibitem{CodelloZanusso2013} A.~Codello and O.~Zanusso,
\textit{On the non-local heat kernel expansion,}
J. Math. Phys. {\bf 54} (2013) 013513,
%   doi:10.1063/1.4776234
arXiv:1203.2034.

\bibitem{BMZ} A.O.~Barvinsky, Y.V.~Gusev, G.A.~Vilkovisky
and V.V.~Zhytnikov,
\textit{The one loop effective action and trace anomaly in four-dimensions,}
Nucl. Phys.  \textbf{B439} (1995) 561, %% -582
%%   doi:10.1016/0550-3213(94)00585-3
hep-th/9404187;
\\
A.O.~Barvinsky, A.G.~Mirzabekian and V.V.~Zhytnikov,
\textit{Conformal decomposition of the effective action and covariant
curvature expansion,} gr-qc/9510037.

\bibitem{BuGui}
I.L. Buchbinder, G. de Berredo-Peixoto and I.L. Shapiro,
{\it Quantum effects in softly broken gauge theories in curved
spacetimes.} Phys. Lett. {\bf B649} (2007) 454, %% -462,
hep-th/0703189.

\bibitem{bavi85}  A.O. Barvinsky and G.A. Vilkovisky,
{\it The generalized Schwinger-DeWitt technique in gauge theories
and quantum gravity,}
Phys. Repts. {\bf 119} (1985) 1.
%%%    DOI: 10.1016/0370-1573(85)90148-6
\,
%% red
%%\textcolor{red}{
\bibitem{Brown-Collins_1980} L.S.~Brown and J.C.~Collins,
\textit{Dimensional renormalization of scalar field theory
in curved space-time,}
Ann. Phys. \textbf{130} (1980) 215.
%%   doi:10.1016/0003-4916(80)90232-8

%%%  end of red

\bibitem{MotVaul} E. Mottola and R. Vaulin,
\textit{ Macroscopic effects of the quantum trace anomaly,}
Phys. Rev. {\bf D74} (2006) 064004,
%% DOI: 10.1103/PhysRevD.74.064004
gr-qc/0604051.
		
\bibitem{star} A.A. Starobinsky,
{ \it A new type of isotropic cosmological models without
singularity},
Phys. Lett. {\bf B91} (1980) 99.
%% DOI: 10.1016/0370-2693(80)90670-X

\bibitem{star83} A.A.~Starobinsky,
{\it The perturbation spectrum evolving from a nonsingular initially
de-Sitter cosmology and the microwave background anisotropy,}
Sov. Astron. Lett. {\bf 9} (1983) 302.

\bibitem{StabInstab}
T.d.P.~Netto, A.M.~Pelinson, I.L.~Shapiro and A.A.~Starobinsky,
{\it From stable to unstable anomaly-induced inflation,}
Eur. Phys. J. {\bf C76} (2016)  544,
%%  doi:10.1140/epjc/s10052-016-4390-4
arXiv:1509.08882.
		
\bibitem{DW-07} S. Deser and R.P. Woodard,
\textit{ Nonlocal Cosmology},
Phys. Rev. Lett. {\bf 99} (2007) 111301,
%% DOI: 10.1103/PhysRevLett.99.111301
astro-ph/0706.2151.

\bibitem{Sponta} E.V. Gorbar and I.L.~Shapiro,
\textit{ Renormalization group and decoupling in curved space
3. The case of spontaneous symmetry breaking},
JHEP {\bf 02} (2004) 060,
%% DOI: 10.1088/1126-6708/2004/02/060
hep-ph/0311190.

\bibitem{Magg2}
M. Maggiore, L. Hollenstein, M. Jaccard and E. Mitsou,
{\it Early dark energy from zero-point quantum fluctuations.}
Phys. Lett. {\bf B704} (2011) 102,\
%%    DOI : 10.1016/j.physletb.2011.09.010
arXiv:1104.3797.
		
\bibitem{ColeWein} S.R. Coleman and E.J. Weinberg,
{\it Radiative corrections as the origin of spontaneous symmetry
breaking,} Phys. Rev. {\bf D7} (1973) 1888.

\bibitem{BuchWolf} I.L. Buchbinder and J.J. Wolfengaut,
\textit{Renormalization group equations and effective action in
curved space-time,}
Class. Quant. Grav. {\bf 5} (1988) 1127. %% -1136

%%%%%%%%%%%%%   New refs - EPJC

}

\end{thebibliography}
	
\end{document}